\documentclass[aps,prd,reprint,superscriptaddress,showkeys,amsmath,amssymb,amsfonts]{revtex4-2}
\usepackage[T1]{fontenc}
\usepackage[utf8]{inputenc}
\usepackage[spanish,USenglish]{babel}
\usepackage[colorlinks,citecolor=blue,urlcolor=blue,linkcolor=blue]{hyperref}
\usepackage{braket}
\usepackage{mathtools}
\usepackage{enumerate}
\usepackage{dcolumn}
\usepackage{enumerate}
\usepackage{bm}

\DeclarePairedDelimiter{\abs}{\lvert}{\rvert}

\newcommand{\BO}{B\nobreakdash-O}

\begin{document}

\title{Strong decays of the lowest bottomonium hybrid within an extended Born-Oppenheimer framework}
\author{R. Bruschini}
\email{roberto.bruschini@ific.uv.es}
\affiliation{\foreignlanguage{spanish}{Unidad Teórica, Instituto de Física Corpuscular (Universidad de Valencia--CSIC), E-46980 Paterna (Valencia)}, Spain}
\author{P. González}
\email{pedro.gonzalez@uv.es}
\affiliation{\foreignlanguage{spanish}{Unidad Teórica, Instituto de Física Corpuscular (Universidad de Valencia--CSIC), E-46980 Paterna (Valencia)}, Spain}
\affiliation{\foreignlanguage{spanish}{Departamento de Física Teórica, Universidad de Valencia, E-46100 Burjassot (Valencia)}, Spain}


\begin{abstract}
We analyze the decays of the theoretically predicted lowest bottomonium hybrid
$H(1P)$ to open bottom two-meson states. We do it by embedding a quark pair
creation model into the Born-Oppenheimer framework which allows for a
unified, QCD-motivated description of bottomonium hybrids as well as bottomonium. A
new $^{1}\!P_{1}$ decay model for $H(1P)$ comes out. The same analysis applied
to bottomonium leads naturally to the well-known $^{3}\!P_{0}$ decay model. We
show that $H(1P)$ and the theoretically predicted bottomonium state
$\Upsilon(5S)$, whose calculated masses are close to each other, have very
different widths for such decays. A comparison with data from $\Upsilon
(10860)$, an experimental resonance whose mass is similar to that of
$\Upsilon(5S)$ and $H(1P)$, is carried out. Neither a $\Upsilon(5S)$ nor a
$H(1P)$ assignment can explain the measured decay widths. However, a
$\Upsilon(5S)$-$H(1P)$ mixing may give account of them supporting previous
analyses of dipion decays of $\Upsilon(10860)$ and suggesting a possible
experimental evidence of $H(1P)$.
\end{abstract}

\maketitle

\section{Introduction}

There is nowadays compelling theoretical evidence, from quenched (without
light quarks) lattice QCD calculations, of the existence of quarkonium hybrids
($Q\overline{Q}g$ bound states where $Q$ is a heavy quark, $b$ or $c$,
$\overline{Q}$ its antiquark, and $g$ stands for a gluon) \cite{Jug99}. In
contrast, there is not convincing experimental evidence of the existence of
hybrids until now mostly due to the difficulty of identifying unambiguous
distinctive signatures for them. As for any other predicted unstable system
these signatures have to be looked at the decay products. Hence a thorough
theoretical analysis of the dominant and/or exclusive decays of hybrids can be
instrumental in unveiling their presence from data. In this regard the lowest
bottomonium hybrid ($b\overline{b}g$ bound state) that we shall call
henceforth $H(1P)$ (the reason for this notation will be explained later on)
may be an ideal system, in spite of its possible mixing with bottomonium
($b\overline{b}$ bound state), for trying to disentangle these signatures for
several reasons.

First, the mass of the $b$ quark, $M_{b}$, is much larger than the QCD scale,
$\Lambda_{QCD}$. This, added to the assumption of no significant string
breaking effects, supports the Born-Oppenheimer ({\BO})
approximation used for the description of $H(1P)$ in QCD. In this approach,
its mass corresponds to the lowest energy level of a Schr\"{o}dinger equation
for $b\overline{b}$ in an excited flavor-singlet {\BO} potential.
This (deepest) hybrid potential, defined by the energy of an excited state of
the gluon field in the presence of static $b$ and $\overline{b}$ sources, has
been calculated in quenched lattice QCD.

Second, the validity of the {\BO} approximation implies
approximate heavy quark spin symmetry (HQSS) \cite{Braa14} which limits its
possible hadronic transitions.

Third, being the lowest hybrid state it cannot decay to other hybrids.
Moreover, as the deepest hybrid {\BO} potential is smaller than
the sum of the ground-state flavor singlet {\BO} potential and the
mass of a glueball (bound state of $gg$) with the appropriate quantum numbers,
decay to a bottomonium state, which is an energy level of the ground-state
flavor singlet {\BO} potential, plus a glueball is not expected.
Hence, its strong decays are constrained to final states not involving hybrids
or glueballs.

In this article we follow these reasonings to analyze the dominant decays of
$H(1P)$ to open-flavor meson-meson states. As this hybrid and the
$\Upsilon(5S)$ bottomonium state, both with the same quantum numbers
$J^{PC}=1^{--}$, are predicted to have about the same mass in the
{\BO} approximation, we study for comparison similar decays from
$\Upsilon(5S)$. In order to be consistent (and to make meaningful the
comparison) we describe the decays, through pair creation, within 
an extended {\BO} framework. For the bottomonium states this description gives
rise to the well-known $^{3}\!P_{0}$ decay model. For $H(1P)$ a $^{1}\!P_{1}$
decay model comes out consistently. It is worth to emphasize that this
$^{1}\!P_{1}$ decay model is essentially different from the decay models built
from constituent glue or flux tube hybrid models, see \cite{Mey15,Far20} and
references therein. In brief, in these hybrid models pair creation is spin
triplet whilst in the $^{1}\!P_{1}$ is spin singlet. This difference, directly
related to the different hybrid descriptions, translates in very distinct
selection rules for the forbidden and allowed decays from $H(1P)$ to open
bottom two-meson states.

These selection rules, and the ones that can be analogously derived for other
hybrids whose disentanglement from data may be even more difficult, could be
of great help for the analysis of upcoming data, in particular from JLab
(GlueX) and Fair (PANDA), or for directing new experimental searches.

The decay width ratios for $\Upsilon(5S)$ and $H(1P)$, calculated from the
$^{3}\!P_{0}$ and $^{1}\!P_{1}$ decay model respectively, are very different.
None of these predicted ratios can explain by its own the measured ones for
the only experimental candidate with a similar mass, the $1^{--}$ resonance
$\Upsilon(10860)$ \cite{PDG20}. In contrast, we show that a $\Upsilon
(5S)$-$H(1P)$ mixed state could give reasonable account of data. This provides
additional support to the mixing scenario proposed in \cite{Bru19} for
$\Upsilon(10860)$ from the study of its leptonic and dipion decays. This
suggest that, altogether, the measured decay widths of $\Upsilon(10860)$ might
provide the first (indirect) experimental evidence of the existence of $H(1P)$.

The contents of this article are organized as follows. In Sec.~\ref{sec2} the
{\BO} framework for the description of bottomonium and bottomonium
hybrids is briefly revisited. In Sec.~\ref{sec3} their dominant strong decays
to open bottom two-meson states are described within an extended {\BO}
framework. Next, in Sec.~\ref{sec4} we apply this description to the
calculation of the widths of the lowest bottomonium hybrid $H(1P)$ and the
$\Upsilon(5S)$ bottomonium state. The results obtained are compared with data
from $\Upsilon(10860)$ in Sec.~\ref{sec5}. This comparison suggests that
$\Upsilon(10860)$ could be a mixing of these two components. Finally, in
Sec.~\ref{sec6}, our main conclusions are summarized.

\section{\label{sec2}Born-Oppenheimer approximation}

The {\BO} approximation, initially developed for the description of molecules
from electromagnetic interactions \cite{BO27} has been shown to be also well
suited for the description of heavy-quark meson bound states from strong
interactions \cite{Jug99,Braa14}. The reason is that it allows for the
implementation of the strong interaction theory (QCD) dynamics through
quenched (with quarks and gluons but no light quarks) lattice QCD
calculations. (For a connection of the {\BO} approximation with effective
field theory, see \cite{Bra18})

%\bigskip

To understand how the {\BO} approximation works let us briefly recall the main
steps in its construction. For this purpose let us consider a meson system
containing a heavy quark-antiquark ($Q\overline{Q}$) interacting with light
fields (gluons), with Hamiltonian
\begin{equation}
H=K_{Q\overline{Q}}+H_{Q\overline{Q}}^{\text{lf}}
\end{equation}
where $K_{Q\overline{Q}}$ is the $Q\overline{Q}$ kinetic energy operator and
$H_{Q\overline{Q}}^{\textup{lf}}$ includes the light field energy operator and
the $Q\overline{Q}$ -- light-field interaction. A bound state $\ket{\psi}$ is
a solution of the characteristic equation
\begin{equation}
H\ket{\psi}=E\ket{\psi}
\end{equation}
where $E$ is the energy of the state. Notice that $\ket{\psi}$ contains
information on both the $Q\overline{Q}$ and light fields.

The first step in building the {\BO} approximation consists in solving the
dynamics of the light fields by neglecting the $Q\overline{Q}$ motion,
i.e., setting the kinetic energy term $K_{Q\overline{Q}}$ equal to zero
(static limit). This corresponds to the limit where $Q$ and $\overline{Q}$ are
infinitely massive, what can be justified because the $Q$ and $\overline{Q}$
masses, $m_{Q}$ and $m_{\overline{Q}}$, are much bigger than the QCD scale
$\Lambda_\textup{QCD}$, which is the energy scale associated with the light fields.
In this limit the quark-antiquark relative position $\bm{r}$ is fixed, ceasing
to be a dynamical variable. Then, in the center of mass reference frame
$H_{Q\overline{Q}}^\textup{lf}$ depends operationally only on the light fields, and
parametrically on $\bm{r}$. We indicate this renaming it as $H_\textup{static}^\textup{lf}(\bm{r})$.

The dynamics of the light fields for any fixed value of $\bm{r}$ can then be
solved from
\begin{equation}
(H_{\textup{static}}^{\textup{lf}}(\bm{r})-V_{\alpha}(\bm{r}))\ket{\alpha;(\bm{r})}=0
\label{STATIC}
\end{equation}
where $\ket{\alpha;(\bm{r})}$ are the light field eigenstates that depend
parametrically on $\bm{r}$ and are characterized by the quantum numbers
$\alpha\equiv(\Lambda,\eta,\epsilon)$. These quantum numbers have been
detailed elsewhere, see for example \cite{Braa14}. Thus, $\Lambda$ stands for
the modulus of the eigenvalue of $\widehat{\bm{r}}\cdot\bm{J}_{\textup{lf}}$,
being $\bm{J}_{\textup{lf}}$ the total angular momentum of the light fields,
$\eta$ for the eigenvalue of $C_\textup{lf}P_\textup{lf}$, being $C_{\textup{lf}}$ the
light field charge-conjugation operator and $P_{\textup{lf}}$ the parity
operator spatially inverting the light fields through the midpoint between $Q$
and $\overline{Q}$, and $\epsilon$ for the eigenvalue of the operator
reflecting the light fields through a plane containing $Q$ and $\overline{Q}$.

As for the eigenvalues $V_{\alpha}(\bm{r})$, depending parametrically on
$\bm{r}$, they correspond to the energies of stationary states of the light
fields in the presence of static $Q$ and $\overline{Q}$ sources placed at
distance $r$. These eigenvalues have been calculated \textit{ab initio} in
quenched lattice QCD \cite{Jug99}.

So, the ground state of the light fields is usually characterized through
$\Sigma_{g}^{+}$ where $\Sigma$ stands for $\Lambda=0$, the subscript $g$ for $\eta=+1$ and
the superscript $+$ for $\epsilon=+1$, and the corresponding eigenvalue reads $V_{\Sigma_{g}^{+}}(\bm{{r}})$.
Up to spin dependent terms that we shall not consider
this eigenvalue mimics the form of the phenomenological Cornell (central)
potential, see for example \cite{Bal01}. As for the first excited state of the
light fields, it is denoted through $\Pi_{u}^{+}$ where $\Pi$ stands for
$\Lambda=1$ and the subscript $u$ for $\eta=-1$, and its corresponding eigenvalue by
$V_{\Pi_{u}^{+}}(\bm{r})$ which, up to spin dependent terms, is also a central potential.

%\bigskip

The second step in building the {\BO} approximation consists in reintroducing
the $Q\overline{Q}$ motion and assuming that the light fields respond almost
instantaneously to the motion of the quark and antiquark.
This is the adiabatic approximation, in
which non adiabatic coupling terms in the kinetic energy are neglected.
Then the bound state equation factorizes in a set of decoupled single-channel
Schr\"{o}dinger like equations for $Q\overline{Q}$, one per each light field
eigenstate $\ket{\alpha; (\bm{r})}$, where the potential is given by the corresponding eigenvalue
$V_{\alpha}(\bm{r})$. The bound state solutions $\ket{\psi}$ can be characterized
as \cite{Braa14}
\begin{multline}
\ket{E,L,m_{L},s_{Q\overline{Q}},m_{s_{Q\overline{Q}}};\alpha }\\
=\int\mathrm{d}{\bm{r}^{\prime}}R_{nL}(r^{\prime})Y_{Lm_{L}}(\widehat{\bm{r}}^{\prime})\ket{\bm{r}^{\prime }}\ket{s_{Q\overline{Q}},m_{s_{Q\overline{Q}}}}\ket{\alpha;(\bm{r}^{\prime})}
\end{multline}
where $L$ and $m_{L}$ indicate that they are eigenstates of $\bm{L}^{2}$ and
$\bm{L}_{z}$, being $\bm{L}$ an angular momentum of the system defined as
$\bm{L}=\bm{l}_{Q\overline{Q}}+\bm{J}_{\alpha}$ where $\bm{l}_{Q\overline{Q}}$
is the orbital angular momentum of $Q\overline{Q}$ and $\bm{J}_{\alpha}$ is the
total angular momentum of the light fields. $s_{Q\overline{Q}}$ and
$m_{s_{Q\overline{Q}}}$ stand for the spin of $Q\overline{Q}$ and its third
component, and $R_{nL}(r)Y_{Lm_{L}}(\widehat{\bm{r}})$ for the wave function
at $\bm{r}$. Notice that the energy of the
state $E$ depends on the quantum numbers $n$ and $L$.

From these solutions, which are also eigenstates of parity with eigenvalue
\begin{equation}
P=\epsilon(-1)^{\Lambda+L+1}
\end{equation}
and charge conjugation with eigenvalue
\begin{equation}
\emph{C}=\eta\epsilon(-1)^{\Lambda+L+s_{Q\overline{Q}}},
\end{equation}
one can easily build physical states
$\ket{E,L,s_{Q\overline{Q}},J,m_{J};\alpha}$ characterized by quantum numbers
$J^{PC}$ where $\bm{J}=\bm{L}+\bm{s}_{Q\overline{Q}}$ is the total angular
momentum of the system.

%\bigskip

For $Q=b$, and the ground state of the light fields
\begin{equation}
\Sigma_{g}^{+}:\, (\Lambda=0,\eta=+1,\epsilon=+1),
\end{equation}
these $J^{PC}$ states correspond to
bottomonium $b\bar{b}$. As $J_{\Sigma_{g}^{+}}=0$ one has
$\bm{L}=\bm{l}_{b\bar{b}}$ and $\bm{J}=\bm{j}_{b\bar{b}}$ where $\bm{j}_{b\bar{b}}$
is the total angular momentum of ${b\bar{b}}$. In particular, for
$J^{PC}=1^{--}$ one has $s_{b\bar{b}}=1$ and $L=l_{b\bar{b}}=0,2$.

%\bigskip

For $Q=b$, and the first excited state of the light fields
\begin{equation}
\Pi_{u}^{+}:\, (\Lambda=1,\eta=-1,\epsilon=+1)
\end{equation}
with $J_{\Pi_{u}^{+}}=1$, the
$J^{PC}$ states correspond to bottomonium hybrids $b\bar{b}g$. The lower
energy state has $J^{PC}=1^{--}$, $s_{b\bar{b}}=0$ and $L=1$. This is why we
use $H(1P)$ to denote this lowest bottomonium hybrid ($H$ for hybrid and $P$
for $L=1)$.

\section{\label{sec3}Strong decays to open bottom two-meson states}

If allowed, the dominant strong decays of bottomonium and bottomonium hybrids
are to open bottom two-meson states. Although the {\BO}
approximation description of bottomonium and bottomonium hybrids incorporates
QCD dynamics through quenched lattice results, the QCD treatment of these
decays requires unquenched (with light quarks) lattice QCD inputs from which
the mixing potentials with open bottom two-meson states can be derived.
These mixing potentials are related to the non-adiabatic coupling terms which
are neglected in the {\BO} approximation. Noticeable progress in the incorporation of 
the non-adiabatic coupling terms within a framework that goes beyond the {\BO} approximation
has been recently reported \cite{Bru20,Bic20}. However, the calculation involving the
non-adiabatic coupling terms requires more unquenched lattice data than currently at disposal \cite{Bal05,Bul19}.
In particular, no lattice data for the mixing of bottomonium hybrids are available. This makes
unaffordable such an \textit{ab initio} treatment at present.

Instead, as the underlying physical mechanism
is light-quark pair creation and string breaking,
we shall assume that the decay proceeds first through light
quark-antiquark ($q\overline{q}$)
pair creation in a transition from an
initial light field (gluon) {\BO} configuration to a light field
(gluon and light quark) configuration given by the direct product of a
{\BO} one and a $q\overline{q}$ state. We call this product an extended {\BO} configuration.

The application of general
conservation arguments to this transition informs us of the possible quantum
numbers of the emitted pair. Then, the recombination of $q\overline{q}$ with
$b\bar{b}$ gives rise to an open bottom two-meson state (string breaking). Although this
two-step process is only an approximation to the QCD mixing, it seems
reasonable to think that, as this mixing takes place via pair creation, the
quantum numbers of the pair will be the same that we have derived from general
conservation laws. These quantum numbers define the decay model, as we show in
what follows.

\subsection{Bottomonium}

For bottomonium it is natural to assume, because of its $b\overline{b}$ content, that
a flavor and color singlet $q\overline{q}$ is emitted within
the hadronic medium. In the extended
{\BO} framework this emission
corresponds to a transition
\begin{multline}
\ket{E,l_{b\bar{b}},s_{b\bar{b}},J=j_{b\bar{b}},m_{J}=m_{j_{b\bar{b}}};\Sigma_{g}^{+}} \\
\rightarrow \ket{E,(l_{b\bar{b}}^{\prime},s_{b\bar{b}},j_{b\overline{b}}^{\prime}),J,m_{J};
(l_{q\overline{q}},s_{q\overline{q}},j_{q\overline{q}}),\Sigma_{g}^{+}}
\end{multline}
where we have used HQSS so that $s_{b\bar{b}}$ is conserved.

Conservation of parity implies
\begin{multline}
\epsilon_{\Sigma_{g}^{+}}(-1)^{\Lambda_{\Sigma_{g}^{+}}+l_{b\bar{b}}+1}
=\epsilon_{\Sigma_{g}^{+}}(-1)^{\Lambda_{\Sigma_{g}^{+}}+l_{b\bar{b}}^{\prime}+1}(-1)^{l_{q\overline{q}}+1}\\
\Rightarrow(-1)^{l_{b\bar{b}}+1}=(-1)^{l_{b\bar{b}}^{\prime}+l_{q\overline{q}}}
\label{CP1}
\end{multline}
and conservation of charge conjugation
\begin{multline}
\eta_{\Sigma_{g}^{+}}\epsilon_{\Sigma_{g}^{+}}(-1)^{l_{b\bar{b}}+s_{b\bar{b}}}
=\eta_{\Sigma_{g}^{+}}\epsilon_{\Sigma_{g}^{+}}(-1)^{l_{b\bar{b}}^{\prime}
+s_{b\bar{b}}}(-1)^{l_{q\overline{q}}+s_{q\overline{q}}}\\
\Rightarrow(-1)^{l_{b\bar{b}}}=
(-1)^{l_{b\bar{b}}^{\prime}}(-1)^{l_{q\overline{q}}+s_{q\overline{q}}} .
\label{CC1}
\end{multline}
By substituting \eqref{CP1} in \eqref{CC1} we have
$(-1)^{s_{q\overline{q}}}=-1\Rightarrow s_{q\overline{q}}=\text{odd}$. Hence
\begin{equation}
s_{q\overline{q}}=1.
\end{equation}
If we reasonably assume that the most favored emission is for $j_{q\overline{q}}$
having its minimal value$,j_{q\overline{q}}=0$, then $l_{q\overline{q}}=1$
(and $l_{b\bar{b}}^{\prime}=l_{b\bar{b}})$, so that the emitted
$q\overline{q}$ pair is in a $^{3}\!P_{0}$ or $0^{++}$ state.

Then, the recombination of the color singlet $q\overline{q}$ with the color
singlet $b\bar{b}$ gives rise to the final $(b\overline{q})$ and
$(\overline{b}q)$ mesons.

This two step process defines the decay model for bottomonium within the
extended {\BO} framework. Actually it corresponds to the so called
$^{3}\!P_{0}$ decay model proposed long time ago \cite{Mic69,LeY73} within a
constituent quark model framework. The $^{3}\!P_{0}$ model, detailed for
bottomonium decays in \cite{Ono}, has been extensively applied in
quarkonium (bottomonium and charmonium) decays to open flavor two-meson states.

This correspondence with the $^{3}\!P_{0}$ decay model is directly related to
the equivalence of the
{\BO} and the constituent quark model (with
a Cornell potential) frameworks for the description of bottomonium. The nice
feature of the extended {\BO} framework is that the possible quantum
numbers for the $q\overline{q}$ pair are derived from general conservation
arguments what also makes us confident in their validity despite our
approximated treatment.

%\bigskip

\subsection{Lowest bottomonium hybrid}

For the lowest bottomonium hybrid $H(1P)$ it is natural to assume,
because of its gluon content, that a color octet light quark-antiquark
pair is emitted within the hadronic medium. In the extended {\BO}
framework, the emission corresponds to a transition
\begin{multline}
\ket{E,L=1,s_{b\bar{b}}=0,J=1,m_{J};\Pi_{u}^{+}} \\
\rightarrow\ket{E,(l_{b\bar{b}}^{\prime},s_{b\bar{b}}
=0,j_{b\overline{b}}^{\prime}),J=1,m_{J};(l_{q\overline{q}},s_{q\overline{q}},j_{q\overline{q}}),\Sigma_{g}^{+}}
\end{multline}
Conservation of parity implies
\begin{multline}
\epsilon_{\Pi_{u}^{+}}(-1)^{\Lambda_{\Pi_{u}^{+}}+L+1}=\epsilon_{\Sigma_{g}^{+}}
(-1)^{\Lambda_{\Sigma_{g}^{+}}+l_{b\bar{b}}^{\prime}+1}(-1)^{l_{q\overline{q}}+1}\\
\Rightarrow-1=(-1)^{l_{b\bar{b}}^{\prime}+l_{q\overline{q}}}
\end{multline}
so that $l_{b\bar{b}}^{\prime}+l_{q\overline{q}}=\text{odd}$, and conservation
of charge conjugation
\begin{multline}
\eta_{\Pi_{u}^{+}}\epsilon_{\Pi_{u}^{+}}(-1)^{\Lambda_{\Pi_{u}^{+}}+L+s_{b\bar{b}}}\\
=\eta_{\Sigma_{g}^{+}}\epsilon_{\Sigma_{g}^{+}}(-1)^{\Lambda_{\Sigma_{g}^{+}}
+l_{b\bar{b}}^{\prime}+s_{b\bar{b}}}(-1)^{l_{q\overline{q}}+s_{q\overline{q}}}\\
\Rightarrow-1=(-1)^{l_{b\bar{b}}^{\prime}+l_{q\overline{q}}+s_{q\overline{q}}}
\end{multline}
so that $l_{b\bar{b}}^{\prime}+l_{q\overline{q}}+s_{q\overline{q}}=\text{odd}$, and $s_{q\overline{q}}=\text{even}$. Hence
\begin{equation}
s_{q\overline{q}}=0.
\end{equation}

Then, using $J_{\Sigma_{g}^{+}}=0$, the total angular momentum
conservation reads $\bm{J}=\bm{l}_{b\bar{b}}^{\prime}+\bm{l}_{q\overline{q}}$.
As $\widehat{\bm{r}}\cdot\bm{l}_{b\bar{b}}^{\prime}=0$ because the orbital
angular momentum of $b\bar{b}$ is orthogonal to the separation vector of $b$
and $\bar{b}$, we have
$\widehat{\bm{r}}\cdot\bm{J}=\widehat{\bm{r}}\cdot\bm{l}_{q\overline{q}}$.
On the other hand $\bm{J}=\bm{L}+\bm{s}_{b\bar{b}}=\bm{L}=\bm{l}_{b\bar{b}}+\bm{J}_{\Pi_{u}^{+}}$
so that $\widehat{\bm{r}}\cdot\bm{J}=\widehat{\bm{r}}\cdot\bm{J}_{\Pi_{u}^{+}}$. Therefore
$\widehat{\bm{r}}\cdot\bm{J}_{\Pi_{u}^{+}}=\widehat{\bm{r}}\cdot\bm{l}_{q\overline{q}}$.
Recalling that the modulus of the eigenvalue of $\widehat{\bm{r}}\cdot\bm{J}_{\Pi_{u}^{+}}$
is $\Lambda_{\Pi_{u}^{+}}=1$ we conclude that
$j_{q\overline{q}}=l_{q\overline{q}}\geq\Lambda_{\Pi_{u}^{+}}=1$.
If we reasonably assume that the most favored emission is for
$j_{q\overline{q}}$ having its minimal value then
$j_{q\overline{q}}=1,s_{q\overline{q}}=0$ and $l_{q\overline{q}}=1$ so that the emitted color
octet $q\overline{q}$ pair is in a $^{1}\!P_{1}$ or $1^{+-}$ state. Then,
$l_{b\bar{b}}^{\prime}=\text{even}$. For the lowest hybrid transition it is
quite natural to assign $j_{b\bar{b}}^{\prime}=l_{b\bar{b}}^{\prime}=0$ so
that the color octet $b\bar{b}$ pair is in a $0^{-+}$ state.

Notice that the quantum numbers of the emitted pair $1^{+-}$ are the same of
the ground state gluelump, which is the limit of the gluon configuration
$\Pi_{u}^{+}$ when $r\rightarrow 0$ \cite{Braa14}.

The second step is the recombination of the color octet $q\overline{q}$ with the
color octet $b\bar{b}$ giving rise to $b\overline{q}$ and $\bar{b}q$ mesons.

This two step process defines the $^{1}\!P_{1}$ model for the decay of $H(1P)$
into open bottom two-meson states within the extended {\BO} framework. Let
us remark that the quantum numbers of the emitted pair $1^{+-}$ are different
from the $1^{--}$ used in decay models of hybrids based on
constituent glue, see \cite{Mey15,Far20} and references therein. Otherwise
said, the bottomonium hybrid descriptions provided by the {\BO}
approximation in QCD and the constituent gluon
models are not equivalent. This difference is crucial to establish the
forbidden and allowed decays of $H(1P)$ to open bottom two-meson states, as we
show next.

\section{\label{sec4}Decay widths}

Let us consider the decay $H(1P)\rightarrow C+F$ where $C$ is a $b\overline
{q}$ meson state $\overline{B}_{(s)}^{(\ast)}$, and $F$ is a $\bar{b}q$ meson
state $B_{(s)}^{(\ast)}$. In parallel with the $^{3}\!P_{0}$ decay model for
bottomonium we shall characterize the $q\overline{q}$ emission by a real
constant probability amplitude: $\sqrt{2}\gamma_{1}$ for $u\overline{u}$ or
$d\overline{d}$ and $\sqrt{2}\gamma_{1}^{\prime}$ for $s\overline{s}$ where
the $\sqrt{2}$ is a color normalization factor. Notice that the color matrix
element in the recombination of the emitted $1^{+-}$ color octet $q\overline{q}$
with the $0^{-+}$ color octet $b\bar{b}$ is $1/\sqrt{2}$ so that the total
(emission + recombination) color factor is $\sqrt{2}\frac{1}{\sqrt{2}}=1$ as it
corresponds to the decay of an initial color singlet into final color singlet
states.
To complete the calculation of the amplitude for the recombination
process (of $q\overline{q}$ with the color octet $b\overline{b}$) we
need the radial wave function of the color octet $b\overline{b}$.
We shall
approximate it by that of the hybrid $R_{n=1,L=1}(r)$. This is justified in
the limit $r\rightarrow0$ where the $r$-dependent interaction potential between $b\overline{b}$ and the
gluon field becomes negligible against the centrifugal barrier. As the orbital angular
momentum of $b\overline{b}$ is zero the orbital part of the wave function is that of
the gluon field and the radial part of the wave function is effectively that of the
color octet $b\bar{b}$ configuration. This wave
function approximation holds as long as the $\Pi_{u}^{+}$ configuration
remains close to the gluelump, what may occur up to a distance around $0.5$ fm
\cite{Braa14}.

The calculation of the width follows exactly the same procedure used in the
$^{3}\!P_{0}$ model detailed in \cite{Mic69,LeY73,Ono}. In the rest frame of
$H(1P)$ and for the emission of a $u\overline{u}$ or $d\overline{d}$ pair it
can be expressed as (for the emission of $s\overline{s}$ one should substitute $\gamma$ by $\gamma^\prime$)
\begin{equation}
\Gamma(H(1P)\rightarrow C+F)
=\gamma_{1}^{2}2\pi\frac{E_{C}E_{F}}{M_{H}}k\abs{\mathcal{M}}^{2}
\label{wid}
\end{equation}
where $M_{H}$ is the mass of the hybrid, $E_{C}$ is the energy of the $C$
meson given by $E_{C}=\sqrt{M_{C}^{2}+k^{2}}$ being $k$ the modulus of the
three-momentum of $C$ (or $F$), and
\begin{multline}
\abs{\mathcal{M}}^{2}=\frac{1}{2\pi^{2}}
\abs{\braket{I_C m_{I_C} I_F m_{I_F} \vert I_{b\bar{b}} m_{I_{b\bar{b}}}}}^{2}\!
\begin{bmatrix}
I_{b} & I_{\overline{b}} & I_{b\bar{b}}\\
I_{q} & I_{\overline{q}} & 0\\
I_{C} & I_{F} & I_{b\bar{b}}
\end{bmatrix}
^{2}\\
\begin{bmatrix}
1/2 & 1/2 & s_{b\bar{b}}\\
1/2 & 1/2 & s_{q\bar{q}}\\
s_{C} & s_{F} & s_{CF}
\end{bmatrix}
^{2}
\begin{bmatrix}
l_{b\bar{b}}^{\prime} & s_{b\bar{b}} & j_{b\bar{b}}^{\prime}\\
l_{q\bar{q}} & s_{q\bar{q}} & j_{q\bar{q}}\\
l_{b\bar{b}}^{\prime}+1 & s_{CF} & J_{H}
\end{bmatrix}
^{2}\abs{\mathcal{J}_{+}(k)}^{2}
\label{Amp}
\end{multline}
where $I$ and $m_{I}$ stand for isospin and its third component, $s$ for spin,
$J$ for total angular momentum of the initial state, $l_{b\bar{b}}^{\prime}=0$,
$s_{b\bar{b}}=0$, $j_{b\bar{b}}^{\prime}=0$, $s_{q\bar{q}}=0$,
$l_{q\bar{q}}=1$ and $j_{q\bar{q}}=1$. The square brackets are related to the
$9j$ symbols:
\begin{equation}
\begin{bmatrix}
j_{a} & j_{b} & j_{e}\\
j_{c} & j_{d} & j_{f}\\
j_{g} & j_{h} & j_{i}
\end{bmatrix}
\equiv\sqrt{\hat{\jmath}_{e}\,\hat{\jmath}_{f}\,\hat{\jmath}_{g}\,\hat{\jmath}_{h}}
\begin{Bmatrix}
j_{a} & j_{b} & j_{e}\\
j_{c} & j_{d} & j_{f}\\
j_{g} & j_{h} & j_{i}
\end{Bmatrix}
\end{equation}
with $\hat{\jmath}\equiv2j+1$. The spatial integral $\mathcal{J}_{+}$ is given
by
\begin{equation}
\mathcal{J}_{+}(k)=i^{l_{b\bar{b}}^{\prime}}
\sqrt{\frac{3(l_{b\bar{b}}^{\prime}+1)}{2l_{b\bar{b}}^{\prime}+3}}\mathcal{I}_{+}(k)
\end{equation}
with
\begin{multline}
\mathcal{I}_{+}(k)=\int_{0}^{\infty}r^{2}\mathrm{d}{r}p^{2}\mathrm{d}{p}
u_{C}^{\ast}(p)u_{F}^{\ast}(p) R_{H(1P)}(r)\\
\bigl[pj_{1}(pr)j_{l_{b\bar{b}}^{\prime}+1}(h_{b}kr)+h_{q}kj_{0}(pr)j_{l_{b\bar{b}}^{\prime}}(h_{b}kr)\bigr],
\end{multline}
where $h_{b}\equiv\frac{M_{b}}{M_{q}+M_{b}}$, $h_{q}\equiv\frac{M_{q}}{M_{q}+M_{b}}$,
and $u$ stands for the Fourier transform of the radial wave function.

Notice that from these expressions one can easily recover the corresponding
ones to the $^{3}\!P_{0}$ model for the decay of an $S$-wave $1^{--}$
bottomonium state $\Upsilon$ by substituting $\gamma_{1}^{2}\rightarrow
\gamma_{0}^{2}$$,s_{q\overline{q}}=0\rightarrow s_{q\overline{q}}=1$,
$j_{q\overline{q}}=1\rightarrow j_{q\overline{q}}=0$,
$s_{b\overline{b}}=0\rightarrow s_{b\overline{b}}=1$,
$j_{b\bar{b}}^{\prime}=0\rightarrow j_{b\bar{b}}^{\prime}=1$, and $H\rightarrow\Upsilon$.

For the sake of simplicity we shall use henceforth the notation:
\begin{equation}
\begin{aligned}
\Gamma_{H_{u}}\equiv\Gamma_{H(1P) \rightarrow B\overline{B}}, & \qquad
\Gamma_{H_{s}}\equiv\Gamma_{H(1P) \rightarrow B_{s}\overline{B}_{s}}, \\
\Gamma_{H_{u}^{\ast}}\equiv\Gamma_{H(1P) \rightarrow B\overline{B}^{\ast}}, & \qquad
\Gamma_{H_{s}^{\ast}}\equiv\Gamma_{H(1P) \rightarrow B_{s}\overline{B}_{s}^{\ast}}, \\
\Gamma_{H_{u}^{\ast\ast}}\equiv\Gamma_{H( 1P) \rightarrow B^{\ast}\overline{B}^{\ast}}, & \qquad
\Gamma_{H_{s}^{\ast\ast}}\equiv \Gamma_{H(1P) \rightarrow B_{s}^{\ast}\overline{B}_{s}^{\ast}}.
\end{aligned}
\end{equation}

From \eqref{Amp} and taking into account that the three elements in the same
column in the $9j$ symbol have to satisfy the triangular rule for the symbol
not to vanish we immediately infer from $s_{q\overline{q}}=0$ that $H(1P)$ can
only decay to $CF$ open bottom two-meson channels with $s_{CF}=s_{b\bar{b}}=0$.
From this spin selection rule, the decays to $B\overline{B}^{\ast}$ and
$B_{s}\overline{B}_{s}^{\ast}$ (we use this notation instead of
$B\overline{B}^{\ast} + B^{\ast}\overline{B}$ and
$B_{s}\overline{B}_{s}^{\ast} + B_{s}^{\ast}\overline{B}_{s}$),
although kinematically allowed, are forbidden
\begin{equation}
\Gamma_{H_{u}^{\ast}}=0\qquad\Gamma_{H_{s}^{\ast}}=0.
\label{WidthsH*}
\end{equation}
As for the calculation of the widths for the other kinematically allowed
decays to $B\overline{B}$, $B^{\ast}\overline{B}^{\ast}$ and
$B_{s}\overline{B}_{s}$, $B_{s}^{\ast}\overline{B}_{s}^{\ast}$ we shall use for
$H(1P)$ the mass $10888$ MeV and the radial wave function calculated in
reference \cite{Bru19}. Let us remind that the excited string potential used in \cite{Bru19}
differs from the lattice QCD parametrization of the hybrid potential \cite{Braa14}
only at short distances, where they are both dominated by the centrifugal
barrier given by $L=1$. Therefore, the difference
between these two parametrizations has no appreciable effect on the wave function of
the lowest bottomonium hybrid. For the final mesons, we use their experimental masses and
for simplicity, as usual, Gaussian radial wave functions with an average rms
radius of $0.45$ fm, as obtained from a Cornell potential model. For the heavy
and light quark masses we have chosen standard values $M_{b}=4793$ MeV,
$M_{s}=500$ MeV and $M_{u,d}=340$ MeV. Thus, we get
\begin{equation}
\begin{aligned}
\frac{\Gamma_{H_{u}}}{\gamma_{1}^{2}} =1.7\text{ MeV}, &\quad
\frac{\Gamma_{H_{u}^{\ast\ast}}}{\gamma_{1}^{2}}=25.7\text{ MeV} , \\
\frac{\Gamma_{H_{s}}}{\gamma_{1}^{\prime2}} =30.3\text{ MeV}, &\quad
\frac{\Gamma_{H_{s}^{\ast\ast}}}{\gamma_{1}^{\prime2}}=114.3\text{ MeV},
\end{aligned}
\label{SWitdthsH}
\end{equation}
so that
\begin{equation}
\frac{\Gamma_{H_{u}^{\ast\ast}}}{\Gamma_{H_{u}}}=15.1,\qquad
\frac{\Gamma_{H_{s}^{\ast\ast}}}{\Gamma_{H_{s}}}=3.8.
\label{RatiosH}
\end{equation}

Let us emphasize again that the hybrid decay pattern resulting from
Eqs.~\eqref{WidthsH*} and \eqref{SWitdthsH} is very different from the one
predicted by constituent glue or flux tube models. In these models, as a
consequence of the assumption of a spin triplet light quark pair,
$s_{q\overline{q}} = 1$, our selection rule \eqref{WidthsH*} does not appear. Instead, a different
selection rule establishing the suppression of the coupling of the $1^{--}$ hybrid with two $S$-wave mesons,
e.g., $B^{(\ast)}\overline{B}^{(\ast)}$, comes out. More concretely, in \cite{Pag97} it has been proved
that when $s_{q\overline{q}}=1$ the amplitude $\mathcal{M}$ in \eqref{wid} for the decay into
two $S$-wave mesons vanishes. As in our case $s_{q\overline{q}}=0$ this selection rule does not apply.

On the other hand, in \cite{Kou05} the authors consider the same decay into two $S$-wave mesons and reach
the same conclusion from their characterization of the hybrid as a
bound state of a $0^{-+}$ $c\overline{c}$ and a gluon in $P$-wave, which is referred
to as a $1^{+-}$ magnetic gluon. This magnetic gluon then decays into a color octet
spin-one $S$-wave light $q\overline{q}$ pair. The fact that
the orbital angular momentum of the gluon in the hybrid is $l_{g}=1$,
determining the symmetry properties of the hybrid wave function entering in the calculation of the amplitude,
is instrumental for the derivation of the selection rule.

A direct comparison of
our {\BO} two-body description of the hybrid with the constituent model three-body description of this reference is not
straightforward, but in this context we can observe that if instead one
considered the hybrid as being made of a $0^{-+}$ $c\overline{c}$
and a $1^{+-}$ \emph{gluelump} in an $S$-wave, with this \emph{gluelump} decaying
into a color octet spin-zero $P$-wave light $q\overline{q}$ pair
as it is our case, then the orbital angular momentum of the \emph{gluelump}
would be zero and the selection rule would not appear.

%\bigskip

It is very illustrative to compare these results with the corresponding decay
widths from $\Upsilon(5S)$ with a calculated mass of $10865$ MeV, quite close
to the hybrid one \cite{Bru19}. In this case, using the $^{3}\!P_{0}$ decay
model and the same kind of self-explained simplified notation we get
\begin{equation}
\begin{aligned}
\frac{\Gamma_{\Upsilon_{u}}}{\gamma_{0}^{2}} &=0.8\text{ MeV},\;
\frac{\Gamma_{\Upsilon_{u}^{\ast\ast}}}{\gamma_{0}^{2}}=0.5\text{ MeV},\;
\frac{\Gamma_{\Upsilon_{u}^{\ast}}}{\gamma_{0}^{2}}=1.9\text{ MeV},\\
\frac{\Gamma_{\Upsilon_{s}}}{\gamma_{0}^{\prime2}} & =0.5\text{ MeV},\;
\frac{\Gamma_{\Upsilon_{s}^{\ast\ast}}}{\gamma_{0}^{\prime2}}=0.3\text{ MeV},\;
\frac{\Gamma_{\Upsilon_{s}^{\ast}}}{\gamma_{0}^{\prime2}}=1.4\text{ MeV},
\end{aligned} \label{Widths5s}
\end{equation}
from which
\begin{equation}
\frac{\Gamma_{\Upsilon_{u}^{\ast}}}{\Gamma_{\Upsilon_{u}}}=2.4,\;
\frac{\Gamma_{\Upsilon_{s}^{\ast}}}{\Gamma_{\Upsilon_{s}}}=2.8,\;
\frac{\Gamma_{\Upsilon_{u}^{\ast\ast}}}{\Gamma_{\Upsilon_{u}}}=0.6,\;
\frac{\Gamma_{\Upsilon_{s}^{\ast\ast}}}{\Gamma_{\Upsilon_{s}}}=0.6.
\label{ratios5S}
\end{equation}
The comparison of Eqs.~\eqref{WidthsH*}, \eqref{RatiosH} with
Eq.~\eqref{ratios5S} makes clear the very different decay patterns for $H(1P)$
and $\Upsilon(5S)$. Next we analyze whether these patterns could provide or
not an explanation for current data.

\section{\label{sec5}Comparison with data}

The only experimental data available to possibly check our results come from
$\Upsilon(10860)$ \cite{PDG20}, a $1^{--}$ bottomonium-like resonance produced
in $e^{+}e^{-}$ annihilation at a c.o.m. energy of $10889.9_{-2.6}^{+3.2}$
MeV, pretty close to the calculated masses of the lowest bottomonium hybrid
$H(1P)$ and the $\Upsilon(5S)$ bottomonium state, and not far from the
$B_{s}^{\ast}\overline{B}_{s}^{\ast}$ threshold.

Although the measured leptonic width $\Gamma_{\Upsilon(10860)
\rightarrow e^{+}e^{-}}=0.31 \pm 0.07$ KeV is compatible within the errors with an
assignment of this resonance to a pure $\Upsilon(5S)$ state, this option is
ruled out because $\Upsilon(10860)$ has dipion decays to
$\pi^{+}\pi^{-}h_{b}((1,2)P)$ and $\pi^{+}\pi^{-}\Upsilon((1,2,3)S)$ with a similar
production rate. As $s_{h_{b}}=0$ and $s_{\Upsilon}=1$, HQSS implies that
$\Upsilon(10860)$ must have $s_{b\bar{b}}=0$ and $s_{b\bar{b}}=1$ components.

As for the ratios of decay widths to open bottom two-meson states a comparison
of the measured values (we do not list the ratio
$\frac{\Gamma_{\Upsilon(10860)\rightarrow B_{s}\overline{B}_{s}}}
{\Gamma_{\Upsilon(10860)\rightarrow B_{s}\overline{B}_{s}^{\ast}}}$ because data on
$\Gamma_{\Upsilon(10860)\rightarrow B_{s}\overline{B}_{s}}$ is very uncertain)
\begin{align}
\frac{\Gamma_{\Upsilon(10860)\rightarrow B\overline{B}}}
{\Gamma_{\Upsilon(10860)\rightarrow B\overline{B}^{\ast}}}&=0.40\pm0.12 \label{BB} \\
\frac{\Gamma_{\Upsilon(10860)\rightarrow B^{\ast}\overline{B}^{\ast}}}
{\Gamma_{\Upsilon(10860)\rightarrow B\overline{B}^{\ast}}}&=2.8\pm0.8 \label{BstarBstar}
\end{align}
and
\begin{equation}
\frac{\Gamma_{\Upsilon(10860)\rightarrow B_{s}^{\ast}\overline{B}_{s}^{\ast}}}
{\Gamma_{\Upsilon(10860)\rightarrow B_{s}\overline{B}_{s}^{\ast}}}=13\pm5 \label{BsBsstar}
\end{equation}
to our results for $\Upsilon(5S)$,
\begin{equation}
\frac{\Gamma_{\Upsilon_{u}}}{\Gamma_{\Upsilon_{u}^{\ast}}} = 0.4, \qquad
\frac{\Gamma_{\Upsilon_{u}^{\ast \ast}}}{\Gamma_{\Upsilon_{u}^{\ast}}} = 0.3, \qquad
\frac{\Gamma_{\Upsilon_{s}^{\ast \ast}}}{\Gamma_{\Upsilon_{s}^{\ast}}} = 0.2,
\end{equation}
also supports a
different assignment for $\Upsilon(10860)$. In this respect, different
proposals about its nature, incorporating $s_{b\bar{b}}=0$ and $s_{b\bar{b}}=1$
components, have been developed.

In Ref.~\cite{Vol12}, following HQSS arguments, a mixture of $\Upsilon(5S)$
and a $P$-wave $B_{s}^{\ast}\overline{B}_{s}^{\ast}$ has been suggested.
Taking into account that the decays $B_{s}^{\ast}\overline{B}_{s}^{\ast}
\rightarrow B^{(\ast)}\overline{B}^{(\ast)}$ are suppressed, $\Upsilon
(10860)\rightarrow B^{(\ast)}\overline{B}^{(\ast)}$ should then proceed through the
$\Upsilon(5S)$ component. However, the comparison of our calculated ratio
$\frac{\Gamma_{\Upsilon_{u}^{\ast\ast}}}{\Gamma_{\Upsilon_{u}^{\ast}}}=0.3$ with data
Eq.~\eqref{BstarBstar} seems not to
support that mixing (notice that in \cite{Vol12} the predicted ratios are
different than ours because the spatial integrals have not been taken into account).

In Ref.~\cite{Ose19} an analysis of the decays through the incorporation of meson-meson
components in the description of the $\Upsilon(10860)$ has been carried out.
A very strong disagreement between the calculated ratio
$\frac{\Gamma_{\Upsilon(10860)\rightarrow B_{s}^{\ast}\overline{B}_{s}^{\ast}}}
{\Gamma_{\Upsilon(10860)\rightarrow B_{s}\overline{B}_{s}^{\ast}}}$
and data has been observed.

%\bigskip

As an alternative, in reference \cite{Bru19} it has been proposed that
$\Upsilon(10860)$ could be a mixing of $H(1P)$ and $\Upsilon(5S)$ (for mixing
in nonrelativistic effective field theories see \cite{Onc17,Ber15})%
\begin{equation}
\Upsilon(10860)=\cos\theta\Ket{\Upsilon(5S)}+\sin\theta\Ket{H(1P)},
\end{equation}
where the probability of $H(1P)$ should be at most of a few percent,
$\sin^{2}\theta\lesssim0.1$, in order to get a good description of the
leptonic widths. This proposal allows for an explanation of the observed dipion
decays. Let us examine now whether it could give or not quantitative account
of the observed decays of $\Upsilon(10860)$ to open bottom meson-meson
channels. For this purpose, we shall calculate the decay widths from the ones
of its $\Upsilon(5S)$ and $H(1P)$ components in the form
\begin{equation}
\Gamma_{\Upsilon(10860)\rightarrow B_{(s)}^{(\ast)}\overline{B}_{(s)}^{(\ast
)}}=\left(\cos\theta\sqrt{\Gamma_{\Upsilon_{u,s}^{(\ast)}}}+\sin\theta
\sqrt{\Gamma_{H_{u,s}^{(\ast)}}}\right)^{2}.
\label{widecon}
\end{equation}

As $\Gamma_{H_{u}^{\ast}}=0$, from \eqref{Widths5s} and \eqref{widecon} we
have
\begin{equation}
\Gamma_{\Upsilon(10860) \rightarrow B\overline{B}^{\ast}}=1.9\gamma_{0}^{2}\cos^{2}\theta \text{ MeV}.
\end{equation}
Then from data
\begin{equation}
\Gamma_{\Upsilon(10860)\rightarrow B\overline{B}^{\ast}}=7.0\pm1.3 \text{ MeV}
\end{equation}
we obtain
\begin{equation}
\gamma_{0}^{2}\cos^{2}\theta=3.7\pm0.7,
\end{equation}
where the quoted errors come from data uncertainty only.

If we use this information, together with the calculated widths
\eqref{SWitdthsH} and \eqref{Widths5s}, then compatibility with experimental
data Eqs.~\eqref{BB} and \eqref{BstarBstar}
translates into two independent conditions to be satisfied:
\begin{equation}
\left\{
\begin{aligned}
(0.95\cdot x+0.65)^{2}&=0.40\pm0.12 \\ 
(3.68\cdot x+0.51)^{2}&=2.8\pm0.8
\end{aligned}
\right.
\end{equation}
where $x\equiv \frac{\gamma_{1}}{\gamma_{0}} \tan\theta$.

Each of these conditions admits two solutions. Full consistency with existent
data requires that there exist a pair of overlapping solutions, so that there
is a value of $x$ that reproduces both experimental ratios within the error
bars. The fact that the two closest solutions are
\begin{equation}
\left\{\begin{aligned}
x &= 0.32 \pm 0.06 \\
x &= -0.02 \pm0.10
\end{aligned}\right.
\end{equation}
(errors from data uncertainty only) shows some tension
(around $2 \sigma$) with data. This may be attributed to the approximations we
have followed, among them having neglected the possible momentum dependence of
$\gamma_{0}$ and $\gamma_{1}$.

Following a Bayesian approach, we work under the hypothesis that these two
solutions are nevertheless two independent measurements of the same quantity,
so we identify the best fit of $x$ with the weighted average:
\begin{equation}
\frac{\gamma_{1}}{\gamma_{0}}\tan\theta=0.23\pm0.14
\end{equation}
where the weighted average error has been doubled in order to account for the
aforementioned uncertainties.

As for the decays to bottom-strange mesons, we use $\Gamma_{H_{s}^{\ast}}=0$
and
\begin{equation}
\Gamma_{\Upsilon(10860)\rightarrow B_{s}\overline{B}_{s}^{\ast}}=
1.4\gamma_{0}^{\prime2}\cos^{2}\theta=0.7\pm0.3\text{  MeV}
\end{equation}
to obtain
\begin{equation}
\gamma_{0}^{\prime2}\cos^{2}\theta=0.5\pm0.2.
\end{equation}
Then from the experimental
ratio Eq.~\eqref{BsBsstar} we get
\begin{equation}
\frac{\gamma_{1}^{\prime}}{\gamma_{0}^{\prime}}\tan\theta=0.35\pm0.16
\end{equation}
Like before, we have doubled the error bar to account for uncertainty coming
from our approximations.

%This is so independently on the amount of $\Upsilon(5S)$-$H(1P)$ mixing. In fact, Eqs.~\eqref{thesol} and \eqref{thesolprime} show that whatever the value of  $\tan\theta$ there are values of $\gamma_1$ and $\gamma'_1$ that give a correct description of $\Upsilon(10860)$ partial widths to open bottom. Hence, independently of whether this resonance contains a minuscule fraction of a very broad hybrid, or a few percent of a narrower hybrid, a fully compatible description of $\Upsilon(10860)$ may come out.

%using \eqref{cosgamma} we get $1.92 \pm 0.18 < \gamma_0 \lesssim 2.03 \pm 1.19$, in good accord with the values commonly used in the literature (see for instance \cite{Ono}), and from \eqref{cosgammaprime} we obtain $0.71 \pm 0.14 < \gamma'_0 \lesssim 0.75 \pm 0.15$. Note how $\gamma'_0$ is smaller than $\gamma_0$, as expected.

Then, for example, for the maximum value of $\sin^{2}\theta=0.1$ we would have
for bottomonium $\gamma_{0}=2.0 \pm 0.2$ in good accord with the value commonly used
in the literature, see for instance \cite{Ono}, and $\gamma_{0}^{\prime}=0.75 \pm 0.15$
so that $\gamma_{0}^{\prime}<\gamma_{0}$ as expected from the more probable
emission of a $u\overline{u}$ or $d\overline{d}$ pair than a $s\overline{s}$
pair. From these values, one would have $\frac{\gamma_{1}}{\gamma_{0}}=0.7\pm0.4$ and
$\frac{\gamma_{1}^{\prime}}{\gamma_{0}^{\prime}}=1.0 \pm 0.5$, so that
for $H(1P)$ we would get $\gamma_{1}=1.4 \pm 0.9$ and
$\gamma_{1}^{\prime}= 0.8 \pm 0.5$.

These results show that a fully consistent explanation of $\Upsilon(10860)$ as
mainly being a mixing of $\Upsilon(5S)$ and the lowest bottomonium hybrid is
feasible. It can be easily inferred that this mixing would be also necessary
to explain data if an additional $\Upsilon(5S)$-$\Upsilon(4D)$ mixing were
implemented, since the decays to $B^{\ast}\overline{B}^{\ast}$ and $B_{s}^{\ast
}\overline{B}_{s}^{\ast}$ from $\Upsilon(4D)$ are even more suppressed than from
$\Upsilon(5S)$.

Therefore, a $\Upsilon(5S)$-$H(1P)$ mixing may give reasonable account of the
observed leptonic, dipion and open-bottom two-meson decays of $\Upsilon
(10860)$. We may tentatively interpret this as an indirect experimental
evidence of the existence of the lowest bottomonium hybrid $H(1P)$.

%\bigskip

A remaining question is whether a direct detection of the hybrid dominated
orthogonal combination, that we shall call $H(10860)$:
\begin{equation}
H(10860)=-\sin\theta\Ket{\Upsilon(5S)}+\cos\theta\Ket{H(1P)}
\end{equation}
is feasible or not. From $0<\sin^{2}\theta\lesssim0.1$ and our previous
results we can easily evaluate a lower limit on the total width of $H(10860)$
to two open-bottom mesons:
\begin{equation}
\Gamma(H(10860)\rightarrow B_{(s)}^{(\ast)}\overline{B}_{(s)}^{(\ast)})\gtrsim300\text{ MeV}.
\end{equation}
This width of at least some hundreds of MeV, and an expected larger width for
the $2P$ hybrid state with a calculated mass around $11080$ MeV, makes very
unlikely a clean experimental signature of $H(10860)$ in the foreseeable future.

%\bigskip

\section{\label{sec6}Summary}

A thorough study of the decays of the lowest bottomonium hybrid, that we have
called $H(1P)$, to open bottom two-meson channels has been carried out by
implementing light-quark pair creation within an extended {\BO} framework.
The application of conservation laws for strong interactions fixates the possible
quantum numbers of the pair. Thus, a $^{1}\!P_{1}$ decay model has been built.
From it selection rules for the forbidden and allowed decays and quantitative
ratios of decay widths have been predicted. These predictions differ greatly
from the ones obtained from the $^{3}\!P_{0}$ decay model resulting for
bottomonium decays within the same framework. In particular, the calculated
ratios for $H(1P)$ and for the bottomonium state $\Upsilon(5S)$, with the same
quantum numbers $J^{PC}=1^{--}$ and predicted masses in the {\BO}
approximation close to each other, have been compared between them and with
the measured widths of $\Upsilon(10860)$, an experimental $1^{--}$ resonance
with similar mass. This comparison indicates that $\Upsilon(10860)$ should not
be assigned to a pure $\Upsilon(5S)$ state,\ in accord with the indications
from Heavy Quark Spin Symmetry when applied to its observed dipion decays. The
need for a heavy-quark spin zero component, apart from the $\Upsilon(5S)$ spin
one, has led to several proposals about the nature of $\Upsilon(10860)$ in the
literature. We have centered on a $\Upsilon(5S)$-$H(1P)$ mixing scenario that
gives reasonable quantitative account of the dipion decay widths. We have
shown that such a mixing could also explain the observed decays of
$\Upsilon(10860)$ to open bottom two-meson channels.
It is worth to emphasize that many of our results,
namely the selection rules and the impossibility
to describe the $\Upsilon(10860)$ as a pure bottomonium
or bottomonium hybrid state, are not affected by any change
in the hybrid wave function. These results
make us tentatively conclude that current data on $\Upsilon(10860)$ may be
showing the presence of the lowest bottomonium hybrid.

\begin{acknowledgments}
This work has been supported by MINECO of Spain and EU Feder Grant No.~FPA2016-77177-C2-1-P, by SEV-2014-0398,
by EU Horizon 2020 Grant No.~824093 (STRONG-2020) and by PID2019-105439GB-C21. R.~B. acknowledges a FPI
fellowship from MICIU of Spain under Grant No.~BES-2017-079860.
\end{acknowledgments}

\bibliography{hybridbib}

\end{document}